# Space-time duality in polariton dynamics

Authors:

Suheng Xu[1,†], Seunghwi Kim[2,†], Rocco A. Vitalone[1,†], Birui Yang[1], Josh Swann[1], Enrico M. Renzi[2], Yuchen Lin[1], Taketo Handa[3], X.-Y. Zhu[3], James Hone[4], Cory Dean[1], Andrea Cavalleri[5,6], M.M. Fogler[7], Andrew J. Millis[1], Andrea Alù[2,8 *], D. N. Basov[1*]

*Affiliations:*
[1] *Department of Physics, Columbia University, New York, New York 10027, USA*
[2] *Photonics Initiative, Advanced Science Research Center, City University of New York, New York, New York 10031, USA*
[3] *Department of Chemistry, Columbia University, New York, New York 10027, USA*
[4] *Department of Mechanical Engineering, Columbia University, New York, New York 10027, USA*
[5] *Max Planck Institute for the Structure and Dynamics of Matter, 22761 Hamburg, Germany*
[6] *Department of Physics, Clarendon Laboratory, University of Oxford, Oxford OX1 3PU, United Kingdom*
[7] *Department of Physics, University of California at San Diego, La Jolla, CA 92093-0319, USA*
[8] *Physics Program, Graduate Center, City University of New York, New York, New York 10016, USA*
*\*Corresponding authors: Andrea Alù: aalu@gc.cuny.edu, D. N. Basov: db3056@columbia.edu*
*†Co-first authors*

## Abstract

The spatial and temporal dynamics of wave propagation are intertwined. A common manifestation of this duality emerges in the spatial and temporal decay of waves as they propagate through a lossy medium. A complete description of the non-Hermitian wave dynamics in such a lossy system, capturing temporal and spatial decays, necessitates the use of complex-valued frequency and/or wavenumber eigen-values. Here, we demonstrate that the propagation of polaritons – hybrid light-matter quasiparticles – can be broadly controlled in space and time by temporally shaping their photonic excitation. Using time-domain terahertz near-field nanoscopy, we study plasmon polaritons in bilayer graphene at sub-picosecond time scales. Suppressed spatial decay of polaritons is implemented by temporally engineering the excitation waveform. Polaritonic space-time metrology data agree with our dynamic model. Through the experimental realization and visualization of polaritonic space-time duality, we uncover the effects of the spatio-temporal engineering of wave dynamics; these are applicable to acoustic, photonic, plasmonic, and electronic systems.

## Introduction

From the atomic scale to interstellar distances, waves transport energy and information through space and time. Wave propagation is invariably accompanied by loss through material dissipation and other non-Hermitian phenomena that exchange energy with the environment. Recent studies have introduced the concept of virtual gain(*1–6*) to mitigate the effect of material losses in defining light-matter interactions without resorting to an active medium. Virtual gain can be achieved by judiciously shaping the amplitude of an oscillating signal in the time domain in order to engage complex-valued eigen-frequencies of an open system. Virtual gain offers pathways to circumvent the impact of material loss in applications such as super-resolution imaging(*5, 7–9*), light scattering(*3*), polariton propagation(*10*), and molecular sensing(*11*). However, some of the initial demonstrations of these concepts have been achieved using data post-processing through linear combinations of images acquired via excitations at multiple real frequencies(*12*).

Here, we demonstrate that space-time duality concepts enable real-time manipulation of the spatial decay rate of surface polariton modes. By leveraging the space-time duality emerging from excitations in the complex frequency plane, we experimentally observe the compensation of spatial polariton decay in lossy materials. Owing to their strong light-matter interactions and electrostatically tunable properties(*13–17*), propagating polaritons in van der Waals materials constitute a unique platform for exploring spacetime duality in wave propagation. Direct imaging of polariton waves simultaneously in space and time by means of nano-THz metrology(*18, 19*) grants comprehensive access to polariton dynamics, losses, and spectral characteristics.

The polariton dynamics in space-time is governed by the wave function, $E(x,t) = Ae^{-i(\hat{\omega}t - \hat{q}x)}$, where generally $\hat{q} = q_r + iq_i$ is a complex wavenumber and $\hat{\omega} = \omega_r + i\omega_i$ is a complex frequency. In the Hermitian limit (absence of losses), $\hat{\omega}$ and $\hat{q}$ are real quantities. However, in realistic lossy materials, it is necessary to consider complex frequencies and wavenumbers. A customary representation of polariton dispersion can be obtained by plotting the imaginary part of the p-polarized reflection coefficient(*20, 21*), as illustrated in Fig. 1 for graphene residing on a dielectric substrate. For excitations at purely real frequencies, as is commonly the case for scanning near-field optical microscopy using continuous-wave laser excitation, ohmic losses result in spatially decaying polariton waves characterized by complex-valued momenta $\hat{q}$ (blue curves in Fig. 1), where $q_i$ defines the spatial decay rate of the mode at any real-valued frequency. Dually, the impact of material loss can be observed through excitations at real-valued momenta, as in plane-wave excitations of polaritons through prism coupling(*22*), yielding complex-valued frequencies $\hat{\omega}$ with decay rate $\omega_i$.

Motivated by the appeal of polaritonic space-time duality, we theoretically explore polaritonic spatio-temporal wave dynamics. Specifically, we consider excitations with tailored temporal profiles comprising a set of complex frequencies that exhibit exponential amplitude decay in time (red and purple curves in Fig. 1a,b); spectra of these complex-frequency excitations reveal broadened frequency content as the amplitude of the THz field diminishes with time (Fig. 1b). Remarkably, the space-time duality of polaritonic waves implies that these temporally decaying inputs can emulate "gain-like" responses even in passive systems(*1–3*). This apparent gain effect is most prominent when the excitation matches the polariton's complex eigen-frequency, thus fully compensating for the spatial decay of the polaritonic wave and yielding polaritons with purely real momenta(*4*, *9*), as illustrated by the red traces in Fig. 1d and Fig. 1e.

We experimentally demonstrate space-time duality by exciting plasmon polaritons with picosecond THz pulses (Fig. 2). Polaritons generated by these terahertz fields exhibit picosecond scale oscillations with a temporally decaying amplitude. We observe spatially sustained polaritonic states whose amplitude shows no spatial decay throughout the duration of the THz field, despite substantial ohmic losses in the material. Utilizing time-domain THz near-field nanoscopy(*23–25*), we map the space-time dynamics of plasmon polaritons in graphene -- a tunable polaritonic medium(*26*, *27*) -- over a broad range of charge carrier densities and temperatures. In our experiments, we visualize sustained spatio-temporal polaritonic propagation that uncovers the roles of causality, space-time duality, and nonlocality in polariton dynamics. When a broadband THz pulse with a temporal decay that defines its complex frequency launches successive oscillations into the polaritonic medium, the pulse also encodes its temporal structure nonlocally across space-time in the polaritonic pattern. The net result is the formation of a polaritonic wave that displays a sustained amplitude in space as it propagates through a lossy medium. We term this phenomenon spatio-temporal virtual gain and demonstrate that it is rooted in space-time duality. Our in-operando implementation inherently incorporates complex frequencies, eliminating the need for post-processing(*10*). Our analytical and numerical models, informed by wave dynamics frameworks[10], reproduce the observed spatio-temporal behavior. These findings establish spatio-temporal virtual gain as a new method for sustaining wave propagation in lossy media.

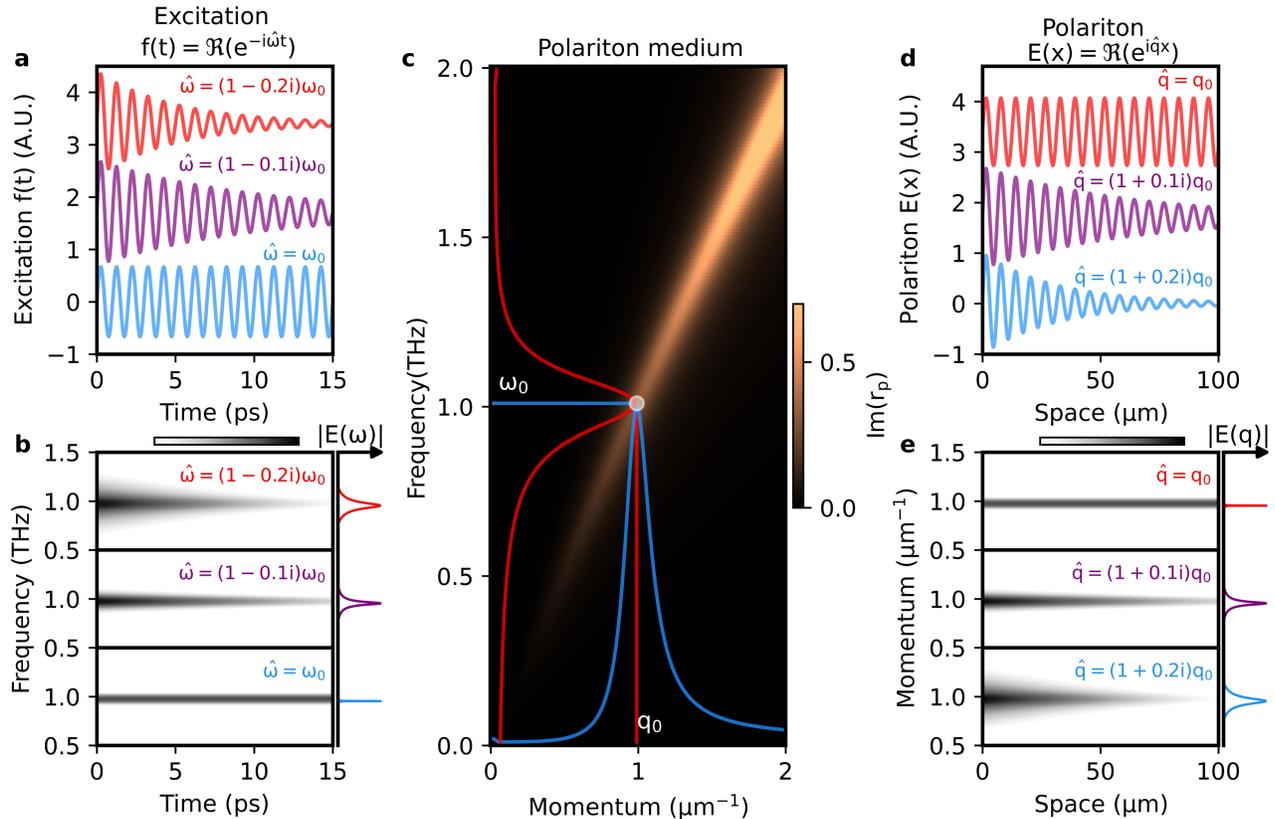

**Figure 1. Complex-frequency excitation of plasmon polaritons and space-time duality.** (a) Time-domain representations of complex-frequency excitations. (b) Frequency content of the excitation pulses in panel a. The false color map shows how the frequency content evolves over time. The frequency spectra at t=0 are displayed on the right. The THz field amplitude decays over time in panel a, corresponding to a broadening of its frequency spectra shown in panel b. (c) Dispersion of plasmon polaritons in a graphene/SiO$_2$ heterostructure. The blue and red curves illustrate the momentum and frequency distributions of excited polaritons. The blue curve corresponds to continuous-wave excitation, while the red curve represents complex-frequency excitation. (d) Polaritonic profiles under the complex-frequency excitations shown in (a). (e) Momentum-space representation of polaritons excited by the corresponding excitations shown in (a). The false color map shows how the momentum component evolves over space. Panels (d) and (e) illustrate spacetime duality features of polaritons, where temporal decay in the excitation pulse compensates spatial decay of the polaritonic wave. Specifically, as the excitation pulse broadens in frequency, the momentum distribution of the launched polaritons narrows, highlighting the dual relationship between the spatial and temporal domains.

## Space-time dynamics of polaritons under pico-second complex-frequency excitation

We explore propagating surface plasmon polaritons (SPPs) in Bernal bilayer graphene with sub-ps temporal resolution and nanometer spatial precision using a time-varying terahertz (THz) excitation, as depicted in Fig. 2a. We utilize photo-conductive antennas (PCAs) to both generate and detect THz pulses; the multi-cycle time-domain profile of one such pulse is shown in Fig. 2b. This broadband THz excitation, spanning a range from 0.5 THz to 1.5 THz, can be described in terms of complex-frequency oscillations as $(\omega_r + i\omega_i)/2\pi \approx (0.8 - 0.3i) THz$ (Fig. 2b). SPPs are visualized using a home-built scattering scanning near-field optical microscope (THz-SNOM)(28–31), which incorporates recent advances in this technique(21, 23, 24, 32–35) within a cryogenic system (see Supplementary Section 1 for a detailed explanation). In our setup, the metallic tip of an atomic force microscope (AFM) is illuminated with THz pulses. The tip couples light into the near field via dipole-dipole interactions(36), launching radially propagating SPPs on graphene. After traveling to and being reflected by the graphene's physical edge, the polariton returns to the AFM tip, where the local polariton field is outcoupled into the far field and detected by a second PCA in the time domain. A schematic of this process is shown in Fig. 2a. The spatial resolution of the tip-demodulated near-field signal is determined by the AFM tip apex radius (> 50 nm), while the temporal resolution (~40 fs) is governed by the width of the optical gating pulse activating the PCA.

The SPP dynamics are tracked via space-time mapping(18, 19). By repeatedly scanning the AFM tip along a path perpendicular to the graphene edge and measuring the local THz electric field at 25-fs intervals, we generate space-time (x-t) maps of the polariton field, displaying the polariton's electric field amplitude as a function of space and time (Fig. 2c). The left panel of Fig. 2c shows the time-dependent polaritonic field measured at the graphene edge, closely following the temporal profile of the terahertz excitation shown in Fig. 2b. The right panel of Fig. 2c shows how this polaritonic field develops in space and time via an x-t map with the edge positioned at x=0. Propagating SPP wave packets manifest as diagonal worldlines (black arrows, Fig. 2c) whose slopes reflect the group velocity of gate-screened acoustic plasmon polaritons[23].

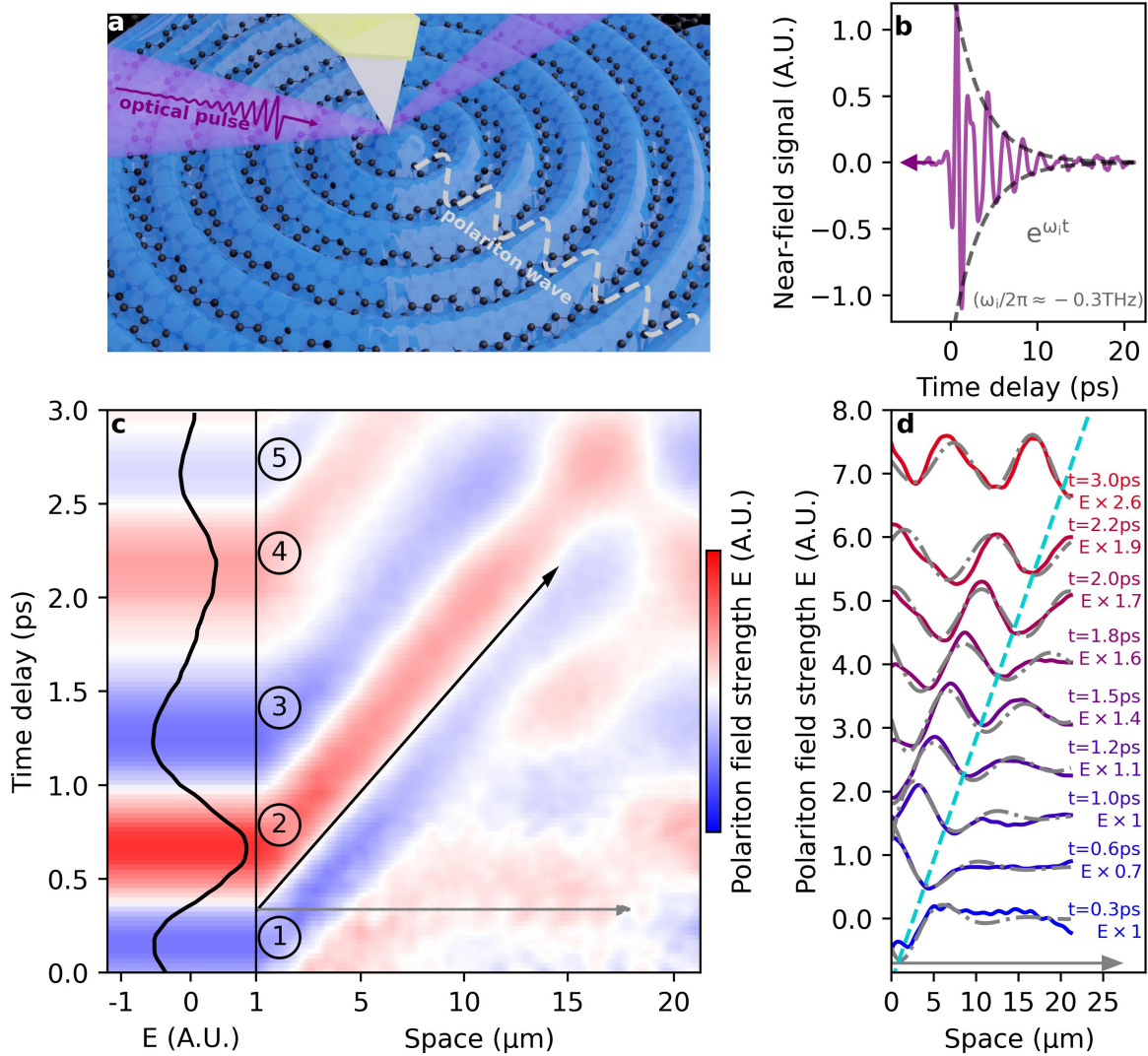

**Figure 2. Polariton space-time dynamics under pico-second terahertz wave excitation.** (a) Schematic illustration of a plasmon polariton excited by time-varying terahertz pulses: a decaying optical pulse (purple) excites both the metallic tip and the sample, initiating spatially sustained plasmon polariton propagation on graphene. (b) Temporal profile of the nano-terahertz pulse collected on the SiO$_2$ substrate. The gray dashed line represents an exponential decay over time, whose decay rate is characterized by the imaginary part of the complex frequency $\omega_i/2\pi = -0.3$THz. (c) Left: Time-dependent polaritonic field measured at the graphene edge (x=0) under pulsed excitation. Right: Space-time $x$-$t$ map of plasmon polariton propagation in bilayer graphene at carrier density $n = 1.13 \times 10^{12}\ cm^{-2}$ and T=38 K. The false color map represents polariton field strength. (d) Solid lines show the spatial profiles of the polaritonic field, extracted along the space axis (indicated by the gray arrow in (c)) at different time delays. Each profile is self-normalized so that the amplitude ranges from -1 to 1, and the field strength at each time delay is multiplied by a factor. Grey dashed lines represent the numerical fits (see Supplementary Section 3 for details). The dashed cyan line marks the polaritonic horizon of the spatially sustained region on each profile. Curves at different time delays are vertically offset for clarity.

# Visualizing sustained polaritonic states under ultrafast complex-frequency excitations

We now explore the spatial profiles of the polariton fields (Fig. 2d): horizontal linecuts of the space-time raw data. In Fig. 2c, we plot these profiles at different time delays. At earlier delays (<1.2 ps), the polariton oscillations are confined near the edge of the structure and decay rapidly with distance. Over time, these oscillations evolve into an extended, wave-like profile with near-constant amplitude: see for example the top trace in Fig. 2d taken at a delay of 3 ps. Below, we demonstrate that this diminished spatial decay is a product of the excitation of polaritons at complex frequencies in our experiments.

Fig. 3 provides a quantitative analysis of the formation of a spatially sustained state observed in the *x-t* map of Fig. 2. In Fig. 3a, we quantify the experimental polariton decay rate by analyzing the spatial field profiles (colored curves) of an individual SPP worldline at different time delays. The peak position shifts away from the edge over time, at a rate governed by the group velocity ($v_g$), while the amplitude decays according to the loss mechanisms. By plotting the polariton field amplitude along the SPP worldline (indicated by the black arrow in Fig. 2c) and fitting its spatial decay with an exponential function, we extract the experimental polariton decay rate ($\Gamma/v_g$) of 0.04 µm⁻¹, which reflects the combined contributions of graphene's electronic scattering and geometric decay(*37*).

In Fig. 3b, we contrast the spatial profiles of a polariton under monochromatic excitations at real frequencies with the results obtained via complex-frequency excitation. The black curve represents the modeled SPP spatial profile under harmonic, constant-amplitude excitation. These monochromatic polaritons typically display a decaying sinusoidal form of the spatial line profile: $E(x) = \cos q_r x \, e^{-q_i x}$, where $q_r = 2\pi/\lambda$ and $q_i = \Gamma/v_g$ are the real and imaginary parts of the SPP wave vector, respectively. In our modeling, we use the parameter $\frac{q_r + i q_i}{2\pi} = (0.35 + 0.04i)\mu m^{-1}$, estimated from the experimental group velocity and polariton decay rate. In stark contrast, the red curve shows the experimentally measured SPP profile at a 3-ps delay under complex-frequency excitation. Notably, excitation at the complex eigen-frequency produces a spatially sustained SPP profile, with a constant oscillation amplitude maintained over a 20-$\mu m$ range. This comparison highlights the pronounced reduction in spatial decay observed when using time-varying and broad spectral range THz pulses to activate plasmon polaritons.

Finally, we analyze the temporal evolution of the spatially sustained SPP profiles. By applying the model $E(x) = \cos q_r x \, e^{-q_i x}$, we extract the complex SPP wave vector from the spatial profiles in the *x-t* map in Fig. 2c as a function of time delay (see Supplementary Section 3). Figs. 3c and 3d show the evolution of $q_r$ and $q_i$ over time. Here, $q_r$ converges around 0.34 $\mu m^{-1}$, indicating a constant SPP wavelength. On the contrary, $q_i$ decreases significantly over time. After approximately 1.2 ps, $q_i$ falls below the experimental polariton decay rate $\Gamma/v_g$ (0.04 µm⁻¹, red dashed line) and approaches zero by 2 ps, signaling the complete suppression of spatial decay and the establishment of fully sustained SPP propagation.

The observed sustained polaritonic states may seem counterintuitive, as graphene's Ohmic loss is independent of either space or time. However, this phenomenon is readily explained by the temporal structure of the exciting THz pulse. Later-launched SPP worldlines (traces 3-5 in Fig. 2c) have lower initial amplitude because the excitation field decreases over time. However, by carefully matching the rate of this temporal decay with the intrinsic polariton loss, we achieve a scenario where later-launched polaritons compensate for the natural decay of earlier-launched ones(traces 1-2 in Fig. 2c). The net effect is a spatially sustained polaritonic state. This progression, culminating in spatially sustained SPP propagation, is demonstrated here through direct measurements for the first time. Subsequent sections detail the theoretical foundation of this phenomenon, which we term polaritonic space-time duality.

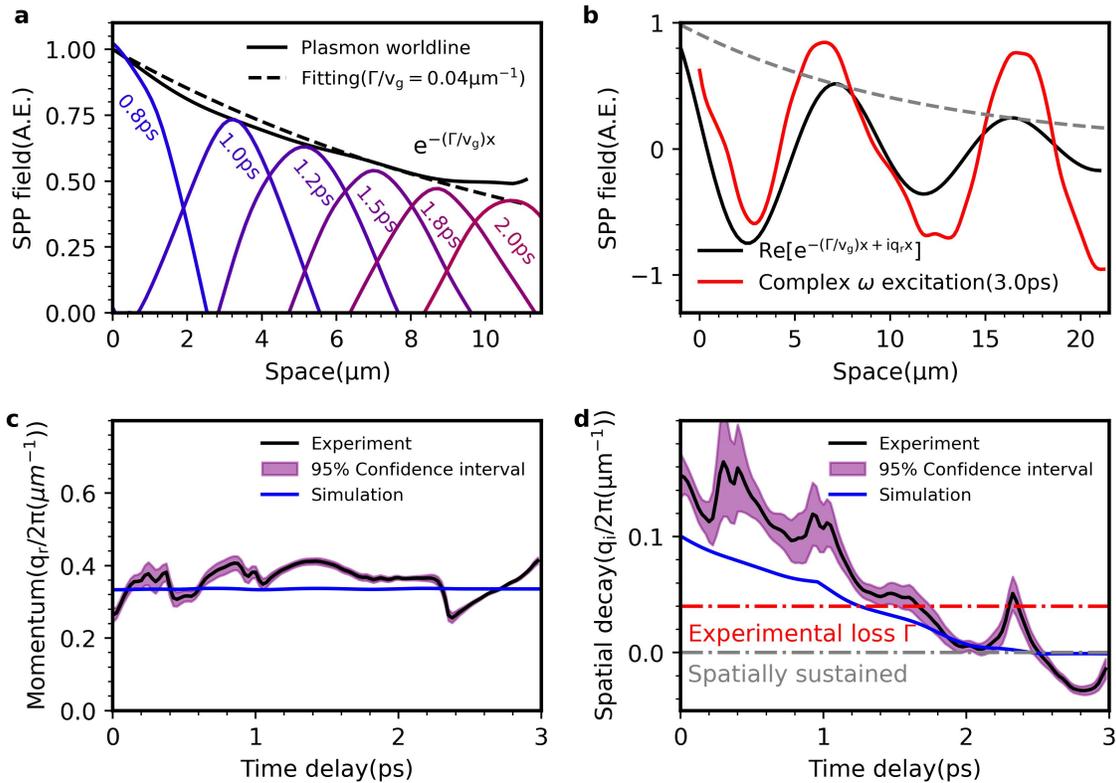

**Figure 3. Compensating polaritonic decay with virtual gain: data and modeling.** (a) Decay of the polariton wave packet along its space-time trajectory. Solid colored lines represent the spatial profiles of the SPP field at different time delays, obtained from the *x-t* map in Fig. 2c. The black solid line represents the polariton field, extracted from a linecut along the worldline: along the direction marked by black arrow in Fig. 2c. The black solid line tracks the polaritonic field amplitude as a function of distance from the sample edge. The decay rate obtained from this polariton field amplitude quantifies the experimental polaritonic dissipation with additive contributions from material loss and geometric decay. (b) Spatially sustained polaritonic profile at a 3-ps time delay. The red curve represents the experimentally observed SPP field profile at a 3-ps delay, whereas the black curve represents the modeled spatial profile of a polaritonic field launched by monochromatic excitation. Under monochromatic excitation, the polaritonic field, characterized by a complex wave vector $\frac{q_r + iq_i}{2\pi} = (0.35 + 0.04i)\mu m^{-1}$, decays with distance

(black). In contrast, the complex-frequency excitation yields the spatially sustained profile (red). (c-d) Time-domain polariton wave vector analysis. Panels (c) and (d) show the extracted real and imaginary parts of the wave vector of the polariton excited by complex-frequency THz pulses as functions of time delay. The black curves represent values extracted from the measured real-space SPP field profiles, while the blue curves represent theoretical predictions evaluated by the custom finite-difference time-domain program (see Supplementary Section 8). The purple shaded areas denote the 95% confidence intervals.

## Space-time duality underlying sustained polaritonic propagation

We now discuss a spatio-temporal dynamic model for interpreting the observed spatially sustained polariton states, capturing the SPP properties under arbitrary excitation and loss conditions. Under monochromatic excitation at real frequencies $f(t) = E_0 \cos \omega_r t$, polaritons with linear dispersion ($\omega_r = v_g q_r$) exhibit spatial decay due to nonzero loss ($\Gamma > 0$), as shown in the $x$-$t$ map in Fig. 4a. These simulation results have been obtained using a custom finite-difference time-domain model as described in Supplementary Sections 4, 5 and 8. The solution of the wave equation $E(x,t) = A \cos(\omega_r t - q_r x) e^{-\frac{\Gamma}{v_g} x}$ confirms exponential spatial decay $\Gamma/v_g$, characteristic of the experimental polariton decay rate.

Conversely, time-varying multicycle pulses with exponentially attenuating amplitudes oscillate at complex frequencies if $f(t) = E_0 e^{\omega_i t} \cos \omega_r t$ where $\omega_i < 0$ for $t > 0$, thus yielding the solution of the wave equation $E(x,t) = A e^{\omega_i t} \cos(\omega_r t - q_r x) e^{-(\omega_i + \Gamma) x / v_g}$ (see Supplementary Section 5). Further, linking complex wave numbers and frequencies, we derive the dispersion relation for lossy polaritons (see Supplementary Section 4,5 for details), whose imaginary component is expressed as

$$\omega_i - v_g q_i = -\Gamma. \tag{1}$$

Full spatial decay compensation ($q_i = 0$) occurs when $\omega_i = -\Gamma$, consistent with the qualitative analysis in reference(10). In this latter regime, we find sustained spatial propagation as depicted in Fig. 4b. Thus, our modeling agrees well with the experimental observations in Fig. 2c and Fig. 2d and with the earlier analysis based on post-processing of the single wavelength polaritonic images(10). The spatially sustained state expands over time, bounded by the polaritonic horizon (cyan dashed line in Fig. 4b); the horizon line reveals the slope $1/v_g$ demanded by causality. Within the horizon, each individual spatial profile maintains a constant oscillation amplitude. However, when comparing successive time slices, the overall amplitude decays uniformly over the entire profile at the polariton decay rate $\Gamma$, reducing to 1/e of its initial value after a time interval of $1/\Gamma$. Beyond the horizon, spatially sustained polariton states cease to exist, complying with the causality of the electromagnetic response.

Under monochromatic excitation, the amplitude of the external optical field remains constant over time, while the polariton field decays spatially. By contrast, complex-frequency excitations enable sustained spatial dynamics. This duality – where decay shifts between space and time – is illustrated in Fig. 4c and Fig. 4d. In a lossy medium, a constant temporal amplitude necessitates spatial decay, while temporal decay allows for spatial stability. In essence, one injects an excess of energy into the polaritonic medium at the earliest times but as time progresses, the energy injection is gradually subdued. This time structure of injection pulses effectively mitigates spatial losses within the polaritonic horizon, as evidenced by sustained propagation.

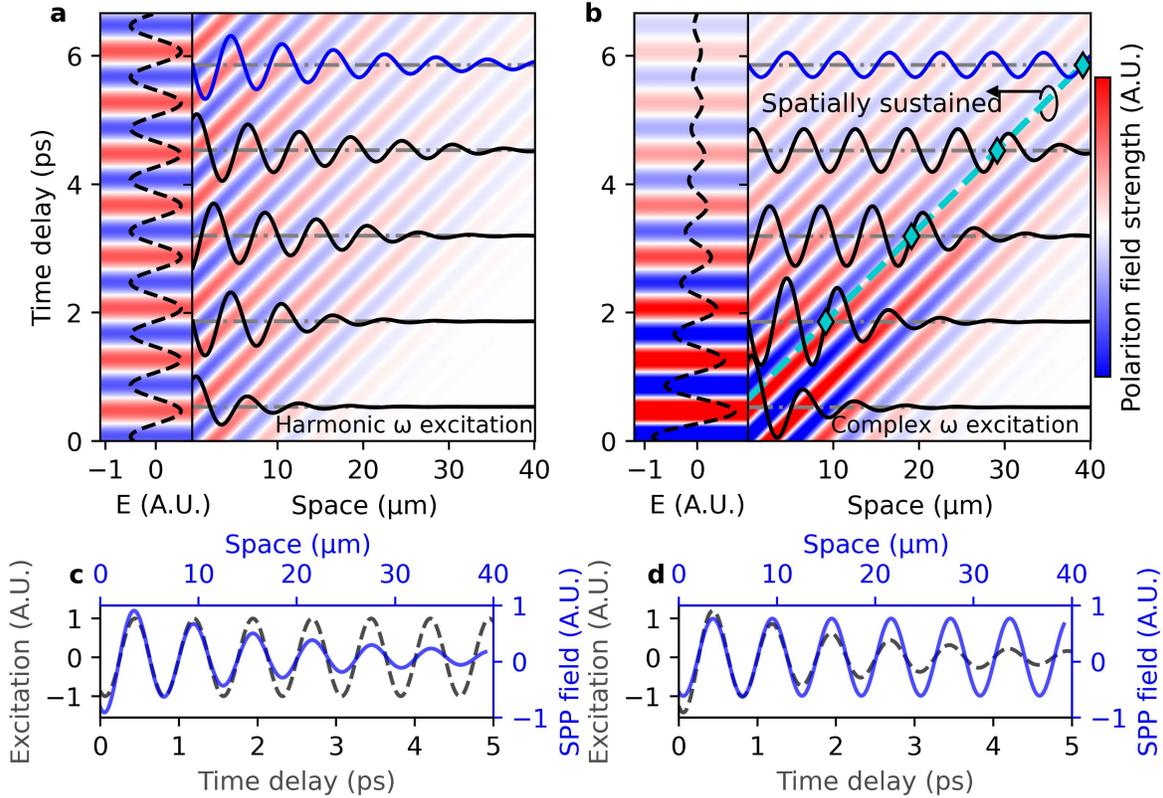

**Figure 4. Numerical simulations of polaritonic losses, sustained propagation and the space-time duality within the polaritonic horizon.** (a) Numerically simulated *x-t* maps of polaritons under monochromatic (harmonic) excitation. Left: Time-dependent polaritonic field at x=0 under the monochromatic excitation. Right: *x-t* map of the polariton, with solid curves representing polariton field spatial profiles at different time delays. (b) Numerically simulated *x-t* maps of polaritons under time-varying excitation. Left: Time-dependent polaritonic field at x=0 under the complex-frequency excitation. Right: *x-t* map of the polariton, where solid curves trace the spatial profiles at various time delays. The purple dots mark the endpoints of the spatially sustained regions on each profile; the dashed purple line highlights the polariton horizon, which defines the overall spatially sustained region in space-time. (c,d) Temporal excitations (0-6 ps) and spatial polariton field profile (at 6 ps) under harmonic (c) and complex-frequency (b) excitations. (c) Under harmonic excitation, the plasmon field strength decays along the space axis. (d) With complex-frequency excitation, where the excitation field decays with time, the polariton field strength remains undiminished along the space axis.

## Outlook: spatio-temporal engineering of polaritonic waves

In Figs. 1-4, we reveal the fundamental interdependence between complex wave numbers and frequencies – space-time duality in the context of polaritons traveling in a medium with ohmic losses. Namely, Eq. (1) quantifies the trade-off between spatial and temporal decay and may even lead to wave amplification. For lossless media ($\Gamma = 0$), Eq. (1) reduces to $q_i = \omega_i/v_g$. Perhaps counterintuitively, Eq. (1) implies that temporally growing excitation ($\omega_i > 0$) induces spatial decay ($q_i > 0$), while temporally decaying excitation ($\omega_i < 0$) enables spatial amplification ($q_i < 0$)(38). In a lossy system ($\Gamma > 0$), a temporally decaying excitation counteracts polariton decay, enabling sustained wave propagation -- a hallmark of space-time duality.

Fig. 5 illustrates the parameter space for spatio-temporal control of polaritonic decay and amplification. The horizontal axis represents the normalized polariton loss ($\Gamma/\omega_r$), while the vertical axis denotes the normalized spatial decay rate ($q_i/q_r$) of propagating waves; both ($\Gamma/\omega_r$) and $q_i/q_r$ are unitless for universal applicability. Diagonal solid lines, predicted by the analytical model in Eq. (1), demonstrate how various temporal structures of the excitation pulse result in readily manipulatable spatial decay. We further verified the outcomes of the analysis in Fig. 5 via numerical simulation (see Supplementary Section 8 for details). We now compare the theoretical and numerical results with the experimental data. The red dots denote the spatial decay rate extracted from experimental SPP *x-t* maps measured at different temperatures and charge carrier densities, where the group velocity and polariton loss vary (see Supplementary Section 9 for a detailed explanation). These data align well with the solid line for the analytical model where $\omega_i/2\pi = -0.3 \, ps^{-1}$, implied by the temporal character of the experimental THz profile. This agreement confirms that our analytical model captures the salient physics underlying our experimental results.

In summary, our experiments demonstrate that time-varying pulsed excitation with natural temporal attenuation alters the spatial decay of polaritons with diverse $\Gamma$. By tailoring temporal decay profiles, we counteract experimental polariton decay, achieving polaritonic waves that propagate with minimal spatial attenuation. This general framework enables the spatio-temporal engineering of polaritonic waves, with implications for photonic, acoustic, and quantum systems. Our findings may be extended to nonlinear regimes and nonlinear dispersion(*13*) (e.g., unscreened plasmons) and suggest pathways to realizing temporal analogs of spatial phenomena, including Floquet engineering, time lenses, and temporal gratings in plasmonic, acoustic, and quantum systems(*39–42*). The ability to directly engineer space-time dynamics via ultrafast pulses opens up avenues for probing exotic wave phenomena inaccessible to traditional methods.

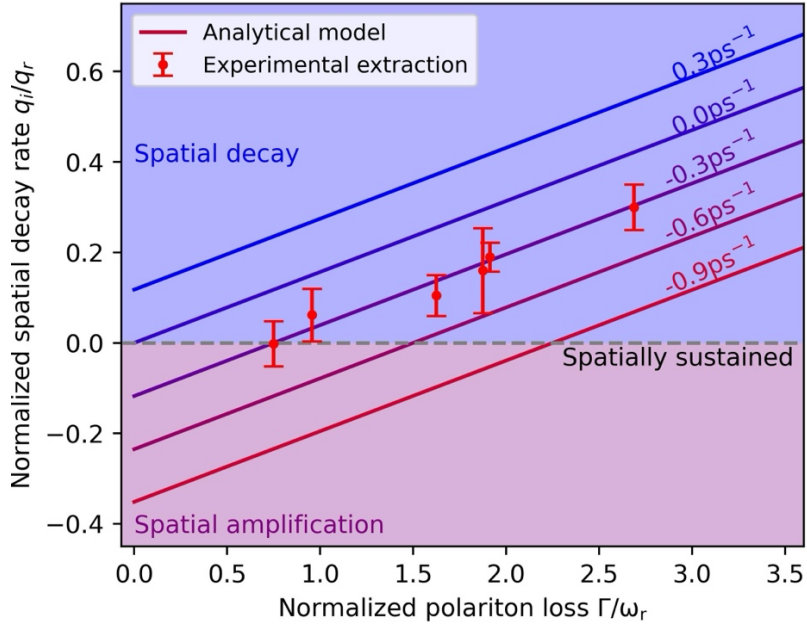

**Figure 5. Space-time engineering of polariton decay.** The map illustrates a universal framework for spatio-temporal control of polaritonic decay. The x-axis represents normalized polaritonic loss, while the y-axis shows the normalized spatial decay rate. Solid colored lines denote analytical predictions of the renormalized spatial decay rate under excitations with varying complex frequency: $\frac{\omega_i}{2\pi} = 0.3, 0.0, -0.3, -0.6, -0.9 \, ps^{-1}$. As the temporal decay rate becomes larger, i.e., a more negative complex frequency $\omega_i$, the accessible parameter space where $q_i < 0$ grows. Red solid dots represent experimental data extracted from *x-t* maps at varying carrier densities or temperatures, with error bars indicating the standard deviation of the wave vector values (a 2-3 ps range). The alignment between analytical, numerical, and experimental results confirms the validity of the spatio-temporal engineering framework.

# Acknowledgement

**Funding:**
Research on THz electrodynamics at Columbia is supported by W911NF2510062 (R.V., D.N.B.). The development of THz instrumentation at Columbia is supported by DOE-BES DE-SC0018426. Fabrication of plasmonic structures, theory and modeling at Columbia are supported as part of Programmable Quantum Materials, an Energy Frontier Research Center funded by the U.S. Department of Energy (DOE), Office of Science, Basic Energy Sciences (BES), under award DE-SC0019443. A.A., S.K., E.M.R. were supported by the Simons Foundation, the Air Force Office of Scientific Research MURI program and the Office of Naval Research.

**Author contributions:**
A.A. and D.N.B. conceived the study. S.X. and R.A.V. recorded the near-field data; S.K. performed theoretical calculations and numerical simulations; B.Y. and J.W prepared the graphene sample with guidance from C.D. and J.H.; S.X., S.K., and R.A.V. analyzed the data with assistance from E.M.R, Y.L., A.C., M.M.F., and A.J.M.; S.X., S.K., R.A.V., A.A., and D.N.B. wrote the manuscript with input from all coauthors.

**Competing interests:**
The authors declare that they have no competing interests.

**Data and materials availability:**
All data needed to evaluate the conclusions in the paper are present in the paper and/or the Supplementary materials.